\documentstyle{article}
\begin{document}
\title{There is no unmet requirement of \\ optical coherence for 
continuous-variable \\ quantum teleportation}
\author{H.M. Wiseman\footnote{H.Wiseman@gu.edu.au. Ph: +61 7 3875 7271. 
Fax: +61 7 3875 7656.} \\ 
Centre for Quantum Dynamics, School of Science, \\
Griffith University,  Brisbane,
Queensland 4111, Australia}
\maketitle

\begin{abstract}
It has been argued [T. Rudolph and B.C. Sanders, Phys. Rev. Lett. {\bf 
87}, 077903 (2001)] that continuous-variable quantum teleportation 
at optical frequencies has not been achieved because the source used 
(a laser) was not `truly coherent'. Here I show that `true coherence'
is always illusory, as the concept of absolute time on a scale beyond 
direct human experience is meaningless. A laser is as good a clock as 
any other, even in principle, and this objection to  teleportation 
experiments is baseless.
\end{abstract}

Recently , Rudolph and Sanders (RS) published a letter entitled 
`Requirement of Optical Coherence for Continuous-Variable Quantum 
Teleportation' \cite{RudSan01}. In it they 
argued that, contrary to  Ref.~\cite{Fur98}, 
continuous-variable quantum teleportation (CVQT) has 
not been and, in fact, cannot be, achieved using a laser as a source
of coherent radiation. They base their argument on their claim 
that a laser is {\em not} a source of coherent radiation, in the 
sense that the output of a laser is not a coherent state, 
but an equal mixture of coherent states with all possible phases.
As they correctly point out, following M\o lmer \cite{Mol96}, 
this can also be interpreted as a 
Poissonian mixture of number states. Thus, they argue, the 
description 
in Ref.~\cite{Fur98} is invalid because it relies upon the 
`partition ensemble fallacy'
\cite{KokBra00}. That is, its analysis is carried using one partition 
of the ensemble  
(into coherent states) 
because it would not be valid in another partition (into number 
states). Although the mathematics of RS is indisputable, 
their letter suffers from a deep conceptual problem.

For the argument of RS to have any teeth, they must allow that 
the production of coherent states of
light is possible in principle, by some means other than a laser.
Indeed they say that `We therefore assert that genuine CVQT 
requires {\em coherent devices}, that is, devices capable of 
generating true coherence, and these are not a feature of current CVQT 
experiments.' If this were true then RS 
would have some basis for criticising at least the formalism of Ref.~\cite{Fur98}. 
It is the point of this letter to show that 
even in principle there are no devices that can generate `true 
coherence' any better than a laser.  Moreover, if one insists on 
rejecting a laser as a coherent source, then one must discard much 
else besides.

Again emphasising that their argument relies upon the possible 
existence of `true coherence', RS say that they `are of course {\em 
not}  asserting that production of coherent states of light is 
impossible: basic quantum electrodynamics
shows that a classical oscillating current can produce coherent
states.' The first problem with this claim is that, as far as we know, the universe is 
quantal, not classical. There is no reason for 
believing in {\em classical} electrical currents.  Nevertheless, 
a suitable quantum current would generate a 
coherent state of light to an arbitrarily good approximation, so I 
will not belabour this point. 

A more 
serious problem is that it is not possible to produce even such a quantum 
current, in a sense that would pass the `truth' test of RS.  
The natural oscillators
at optical frequencies are the electrons in atoms.
Without coherent light, we could still make the atom `ring'  
by `striking' it (with a free electron, for example). 
However, for this oscillation to  have a definite 
phase the time of the collision would have to be known to an
accuracy less than an optical cycle, of order $10^{-15}$s.
Otherwise one would have to average
over all possible phases and one would be left with exactly
the same problem as with the laser. 

Achieving this accuracy would require 
a clock which ticks faster than $10^{15}$s$^{-1}$. Presumably this is 
the sort of `technical challenge' that RS mention in the context of 
producing coherence by making a measurement on the gain medium of a 
laser. But it is much more than just a technical challenge. Instead, 
it just moves the problem back another step. 
How could we be sure that the clock has a definite phase? Perhaps it 
is fixed relative to another clock, but then we can ask the question again of 
that second clock. And so {\em ad infinitum}.

It is thus clear that the `true coherence' as meant by 
RS is impossible to produce technologically. 
It cannot mean coherent relative to any clock, because 
their argument attacking the phase of a laser can be equally used to attack 
that of any clock. Therefore, if it means anything, `true coherence' 
must mean coherent relative to an absolute time standard for the 
universe. Since 
such a hypothetical absolute time standard can never be measured, 
I would maintain that it is meaningless, and with it the idea of `true 
coherence' in the sense of RS. 

 It might be countered that 
as conscious beings we experience the flow of absolute time
directly, and so give it meaning. Admitting  
the validity of this 
temporal experience (which does not go without saying \cite{Peg91}), 
this argument nevertheless cannot work to establish 
optical coherence.  We cannot
simply look at a clock ticking every $10^{-15}$s and verify that it 
has a definite 
phase, because we cannot perceive anything in $10^{-15}$s. 

My rebuttal applies not only to optical frequencies. 
Experiments show that our perception of time
has a resolution in the range of tens or even hundreds of milliseconds 
\cite{Pen90}. Thus {\em we cannot 
establish the absolute phase of any oscillator
of frequency greater than a few tens of Hertz.} At higher frequencies 
we can only establish the phase of one oscillator relative to another oscillator. 
This conclusion is not altered by oscillations obtained by 
frequency combs or $2^{n}$-upling \cite{Jon00}. The timing of the zeros of 
the highest harmonic 
can be no more accurately defined than that of 
the fundamental.

From personal observation, discussions between physicists about the 
existence of `true coherence' or `mean fields' 
often end with one party waving an
arm up and down, intimating that by so moving an electric charge, 
 a mean field would be produced. But that appeal fails precisely 
when the frequency is too fast for us consciously to move 
any part of our body at that frequency. There are many qualitative 
differences between the way radiation is generated, or detected, 
across the spectrum. But the only location for a fundamental 
(if fuzzy) dividing line in frequency between 
absolute and relative phase, or between `true coherence' 
\cite{RudSan01} and `convenient fictions' \cite{Mol96} 
would be between oscillations we can 
observe directly  and those we cannot. If this division is unpalatable 
that is because it relies upon the notion of absolute time.
By abandoning this ill-conceived  notion,  the  
dividing line between supposed `truth' and supposed `fiction' disappears.

With no absolute time, all we can ever do is to use an agreed time 
standard, and measure phase relative to that. 
In this context, a laser beam is as good a 
`clock' as any other. The electrodynamic 
${\bf p}\cdot{\bf A}$ coupling allows, in principle, the laser clock 
to be synchronised with any 
material clock. The latter could then also be synchronised with  
any other clock based on any gauge boson field, through analogous 
coupling Hamiltonians \cite{Ryd85}. The fungibility of their time standards is what 
makes all of these time-keepers equivalent, and justifies calling them 
`clocks'.  This is not an empty definition. An 
atom laser beam is not a clock in this sense. Its phase can only be
defined or measured relative to another atom laser beam of the same 
species \cite{Wis97}.

It might be objected that  a laser beam is not a clock because it cannot 
establish a time standard between arbitrarily many parties. Eventually 
it will run out, or, if it is a continuous beam, its finite linewidth 
will mean that the later part of the beam has a random phase relative 
to the former part. This is of course true, and the fundamental 
limits are set by the 
finiteness of the excitation of the laser mode and laser gain medium 
\cite{Wis99b}. This excitation may be very 
large (measured in units of $\hbar \omega$), but is not infinite. 
However, exactly the same criticism applies to any physical clock, 
even if we are not used to worrying about it for material clocks that 
typically have huge excitations. A similar point has been made in 
Ref.~\cite{EguGarRay99}.

The above arguments lead inevitably to the conclusion that 
in quantum optical experiments 
there is no necessity to consider, even hypothetically, any time-keeper beyond 
the laser which serves as a phase reference.  No other clock is 
superior in any fundamental sense. Now by definition a laser beam is perfectly 
coherent relative to itself (ignoring experimentally negligible phase 
and amplitude fluctuations, and transverse mode incoherence). 
Thus, the phase reference laser beam in the 
teleportation experiment  {\em is} in 
 a coherent state. 
 There is no process that will make a coherent state 
 in any stronger sense, and no need for for any stronger sense. There 
 is no unmet need for optical coherence in continuous-variable 
 quantum teleportation. 

This letter would not be complete without discussing  
the recent papers \cite{EnkFuc02a,EnkFuc02b} by 
van Enk and Fuchs also commenting on Ref.~\cite{RudSan01}.
These can be understood as 
an explicit calculation showing (part of) 
how a laser beam, without an absolute phase, can 
function as a clock; how the phase information can be distributed and how 
there is no harm in regarding the phase as real. This is essentially the 
same point originally made by M\o lmer, that the laser phase is a 
`convenient fiction' \cite{Mol96}. The analysis of van Enk and Fuchs 
gives a rigorous information-theoretic definition of `convenience', in 
terms of the quantum de Finetti theorem \cite{EnkFuc02b}. However, it 
could be questioned whether convenience, however rigorously defined, is 
sufficient for a dispensation from the ban on the `partition 
ensemble fallacy'.

The real lesson that should have 
been drawn from the analysis of RS is 
that quantum teleportation can be, and should have been, defined operationally, 
so that the reality of the laser phase would have been 
irrelevant. The reality of the laser phase can nevertheless be defended, 
as I have done, on the grounds that it is no less real than any other 
phase.  Unfortunately, 
van Enk and Fuchs appear to accept the position of RS, that there is 
such a concept as `true coherence', in that there is a 
time standard more real than that offered 
by the laser itself. Indeed, they say \cite{EnkFuc02a} 
\begin{quote}
However, recent developments
\cite{Jon00} may make it possible to compare the phase of
an optical light beam directly to the phase of a microwave
field. Using this technique, the only further measurement
required \ldots is a measurement of
the {\em absolute phase} [my emphasis] of the microwave field, which is possible
electronically. This measurement would create an
optical coherent state from a standard laser source for the
first time.
\end{quote} 
The implication is that an electronic measurement 
somehow makes the phase real, which it was not when it was an 
optical phase.  As I have argued above, there is no reason to regard 
the laser phase as any less real than any other phase. There is nothing 
gained in, and no need for, appealing to any other clock in order to say that 
the laser is in a coherent state. If a standard laser source is 
used as the clock in an experiment then it creates an 
optical coherent state already. 

To conclude, Rudolph and Sanders' concept of `true coherence' 
requires (and van Enk and Fuchs 
appear to accept) the existence of an absolute time standard more real 
than that offered by the laser oscillation. I have argued that such a time 
standard is illusory. A laser is as good a clock as any other, and 
there is no need to look beyond current lasers to find 
optical coherence. For van Enk and Fuchs coherent states are still 
fictions, but their use is justified 
on the grounds of convenience, defined rigorously using the quantum de 
Finetti theorem. Rudolph and Sanders, by contrast, insist 
that if we cannot measure the phase of an oscillator (such as a 
laser) relative to their assumed absolute time, then we must assign it 
a mixed state, averaged over all possible phases. 
Their arguments, carried to their 
logical conclusion, would banish 
from our theories not only coherent states for lasers, but 
any time $t$ or phase $\phi$ if its 
implied resolution were beyond that of direct human experience. 
To scientists and engineers, this would be unacceptable pedantry.

\section*{Acknowledgments}

I wish to acknowledge recent discussions with T. Rudolph, B. C. Sanders, 
C. Fuchs and S. L. Braunstein.


\end{document}